\documentclass[prb,aps,twocolumn,superscriptaddress]{revtex4-1}
\usepackage{amsmath}
\usepackage{latexsym}
\usepackage[english]{babel}
\usepackage{graphicx}
\DeclareGraphicsExtensions{.pdf,.png,.jpg, .jpeg, .eps}
\usepackage{bm}
\usepackage{array}
\usepackage{amssymb}
\usepackage{amsfonts}
\usepackage{multirow}
\usepackage{epstopdf}
\usepackage{color}
\usepackage{ulem}

\date{\today}

\begin{document}

\title{Giant electromagnetic proximity effect in superconductor/ferromagnet superlattices}
	
	\author{A. V. Putilov}
	\affiliation{Institute for Physics of Microstructures, Russian Academy of Sciences, 603950 Nizhny Novgorod, GSP-105, Russia}
	\author{S. V. Mironov}
	\affiliation{Institute for Physics of Microstructures, Russian Academy of Sciences, 603950 Nizhny Novgorod, GSP-105, Russia}
	\author{A. S. Mel'nikov}
	\affiliation{Institute for Physics of Microstructures, Russian Academy of Sciences, 603950 Nizhny Novgorod, GSP-105, Russia}
	\author{A. I. Buzdin}
	\affiliation{University Bordeaux, LOMA UMR-CNRS 5798, F-33405 Talence Cedex, France}
	\affiliation{World-Class Research Center ``Digital Biodesign and Personalized Healthcare'', Sechenov First Moscow State Medical University, Moscow 119991, Russia}
	
\begin{abstract}
We show that in superlattices with alternating superconducting (S) and ferromagnetic (F) layers the spontaneous magnetic field induced in the superconducting layers due to the electromagnetic proximity effect becomes dramatically enhanced compared to the previously studied S/F bilayers. The effect reveals itself for the in-plane orientation of the magnetic moments both for ferromagnetic and anti-ferromagnetic ordering of the moments in the F layers. In the finite size samples the magnetic field decays from the sample surface towards the bulk of the structure, and the decay length strongly depends on the relative orientation of the sample surface, the layers planes and magnetic moments in the F layers. The obtained results provide additional insights into experimental data on the neutron scattering in Nb/Gd superlattices.   
\end{abstract}
	
\maketitle

\section{Introduction}

The proximity effect in superconductor (S) -- ferromagnet (F) hybrid structures is known to be responsible for rich variety of exciting interference phenomena affecting thermodynamic and transport properties of these systems \cite{Buzdin-RMP-05, Bergeret-RMP-05, Golubov-RMP-04, Eschrig-AdvPhys-06, Linder-NatPhys-15}. The Cooper pairs penetrating the ferromagnet change their spin structure and become spin-polarized under the effect of the exchange field so that the superconducting correlation function contains both spin-singlet and spin-triplet components. This spin transformation results in the non-monotonous dependence of the critical temperature and the in-plane critical current on the F layer thickness in planar S/F structures \cite{Buzdin-JETPL-90, Jiang-PRL-95, Zdravkov-PRL-06, Buzdin-JETP-92}, the formation of Josephson $\pi$-junctions \cite{Buzdin-JETPL-82, Ryazanov-PRL-01}, increase in the electronic density of states at the Fermi level \cite{Buzdin-PRB-00, Kontos-PRL-01, Braude-PRL-07, Cottet-PRL-11}, and so on. At the same time, the F layer affects the superconductor by inducing different types of magnetic ordering there. There are two dominating mechanisms responsible for this back-action. The first one is related to the penetration of the spin-polarized Cooper pairs from the ferromagnet back to the superconductor (the so-called inverse proximity effect) \cite{Bergeret-PRB-03, Bergeret-PRB-04, Bergeret-EPL-04, Bergeret-PRB-05, Volkov-PRB-19, Krivorushko-PRB-02, Lofwander-PRL-05, Faure-PhysC-07, Kharitonov-PRB-06, Salikhov-PRL-09}. The resulting spin polarization in the S layer is localized at the scale of the Cooper pair diffusion length near the S/F interface which has the order of the superconducting coherence length $\xi_s$. The second mechanism originates from the effect of the stray magnetic fields produced by non-uniform magnetization of the ferromagnet revealing through the generation of the Meissner screening currents and vortex structure in the adjacent superconductor (see, e.g., Refs.~\onlinecite{Aladyshkin, Eremin, Milosevic} and Ref.~\onlinecite{Aladyshkin-SuST-09} for a review). Note that for S/F structures with the uniform in-plane magnetization in the F layers both mechanisms predict negligibly small magnetic fields arising in the S subsystem at distances much larger than $\xi_s$ from the S/F interface.

However, the recent theoretical and experimental works\cite{Mironov-APL-18, Devizorova-PRB-19, Flokstra-APL-19, Stewart-PRB-19, Mironov-JETPL-21} unveiled one more unusual consequence of the proximity effect in S/F structures which is responsible for the anomalous enhancement of the stray magnetic fields generated in the superconducting subsystem even for the case of uniform in-plane magnetization in the F layer. This phenomenon called an electromagnetic proximity effect (EPE) is based on the fact that the Cooper pairs penetrating into the F layer interact with the magnetization field $4\pi M$, which results in a formation of Meissner screening current inside the ferromagnet. This current induces an additional magnetic field which penetrates the superconductor and becomes screened at a distance of the order of the London penetration depth $\lambda$ which strongly exceeds the Cooper pair diffusion length $\xi_s$ in type-II superconductors. Such anomalous long-range spread of the magnetic field into the superconducting part of S/F systems should naturally affect the operation regimes of various logical, memory and quantum computing elements of superconducting spintronics \cite{Eschrig-AdvPhys-06}. In these devices the penetration of the stray magnetic field into the superconductor is often considered as an undesirable effect because it leads to the uncontrollable generation of the Meissner currents and possible vortex entrance. On the other hand, the sensitivity of the EPE to the magnetic moment configuration \cite{Devizorova-PRB-19} may provide additional mechanisms for the control of the superconducting properties of the cryogenic S/F devices.

Experimentally EPE reveals itself in an additional long-range  magnetic field arising inside the S layer which can be detected via SQUID magnetometry\cite{Wu-PRB-07, Nagy-EPL-16}, in the polar Kerr effect measurements \cite{Xia-PRL-09}. Even more detailed information about the spatial distribution of the magnetic field in the layered S/F structures can be extracted from the low-energy muon spin-rotation experiments\cite{Flokstra-NatPhys-16, Bernardo-PRX-15, Flokstra-APL-19, Stewart-PRB-19, Flokstra-PRL-18} or neutron scattering measurements\cite{Khaydukov-PRB-19, Khaydukov-PRB-14, Khaydukov-JETPL-13}. In these experiments the temperature decrease below the superconducting critical temperature is accompanied by the generation of the magnetic field outside the F layers penetrating the superconducting subsystem at the length-scale comparable to the London penetration depth and substantially larger than the coherence length. In addition, in S/F$_1$/F$_2$ structures the amplitude of the induced magnetic field was shown to become enhanced for the perpendicular mutual orientation of magnetic moments in the F$_1$ and F$_2$ layers\cite{Flokstra-NatPhys-16} which is in agreement with the theoretical calculations of the EPE in such hybrids\cite{Devizorova-PRB-19}.

Note that the magnitude of the field ${\bf B}$ in S/F bilayers emerging due to EPE appears to be small compared to the magnetization field $4\pi M_0$ inside the ferromagnet ($M_0$ is the magnetization of the F layer). The ratio $B/(4\pi M_0)$ for typical S/F structures has the order of $(\xi_f/\lambda)^2\sim10^{-2}$ ($\xi_f$ is the superconducting coherence length inside the ferromagnet), so that in the systems where the F layer produces the stray magnetic field due to the non-uniform magnetization pattern the contribution from the EPE should be small.

In this paper we show that the electromagnetic proximity effect becomes dramatically enhanced in the superlattices consisting of alternating thin superconducting and ferromagnetic layers as compared to the S/F bilayer system. Such strengthening of the EPE arises from the additive contributions to the magnetic field coming from each F layer, and the incremental growth of the spontaneous field becomes limited only when the total thickness of the superlattice becomes of the order of $\lambda$. As a result, the maximal ratio between the spontaneous field and the magnetization field $4\pi M_0$ reaches the value $d_f/(d_s+d_f)$ where $d_s$ and $d_f$ are the thicknesses of the S and F layers of the superlattice, respectively.  

The physics of the EPE in S/F superlattices with $d_s,~d_f\ll\lambda$ appears to be very similar to the one in the bulk ferromagnetic superconductors with dominating orbital mechanism of magnetic interaction \cite{Bulaevskii-AdvPhys-85}. Deep inside the sample (at distances much larger than $\lambda$ from the faces) the magnetic field averaged over the period of the superlattice should vanish so that the magnetization field produced by F layers is totally compensated by the Meissner currents. As a result, the magnetic field in the S layer is $B_S=-4\pi M_0d_f/(d_s+d_f)$ while inside the F layer it is equal to $B_F=4\pi M_0 d_s/(d_s+d_f)$.
Remarkably, the length scale which provides a transition from the distribution $\mathbf B(\mathbf r)$ near the sample surface  to the bulk values $B_F$ and $B_S$ substantially depends on the relative orientation of the sample surface, the S/F interfaces and the magnetic moment $\mathbf M$ in the F layers. Specifically, this decay length $\lambda_0\sim\sqrt{\left<\lambda^2({\bf r})\right>}$ for the sample surface which is parallel to the ${\bf M}_0$ vector but perpendicular to the S and F layers (here the angular brackets denote the spatial averaging over the period of the S/F superlattice).
Here $\lambda(\mathbf r)$ is the spatial profile of the local London penetration depth which is typically non-uniform for the structures with $d_f$ slightly exceeding the coherence length $\xi_f$ inside the ferromagnet.
At the same time, for the surface parallel to the planes of the superlattice $\lambda_0\sim1/\sqrt{\left<\lambda^{-2}({\bf r})\right>}$.
Our calculation proves that the EPE effect in S/F superlattices is of primary importance, in particular for the adequate interpretation of the experimental data of the neutron scattering measurements.

Finally we demonstrate that the electromagnetic proximity effect in S/F superlattices with the anti-ferromagnetic ordering between the magnetic moment in the neighboring F layers can result in a strong stray field penetrated into a thick superconductor adjacent to the superlattice. Naively, one can expect that the EPE in such structures should be negligibly small since the average magnetization is zero. However, the compensation of the magnetic fields produced by the neighboring F layers appears not to be full, and the resulting magnetic field has the order of $d_f/\lambda$ which is even larger than the magnetic field generated in S/F bilayers. 

Let us emphasize that the contribution of the inverse proximity effect into the distribution of magnetization is not taken into account in our calculations.
This inverse proximity effect is responsible for the spin polarization of electrons and subsequent magnetization of the superconducting surface layer with the width of the order of the Cooper pair size, i.e., the superconducting coherence length $\xi_s$\cite{Volkov-PRB-19}.
The very generic estimate for the magnetization of the S layer induced by the inverse proximity effect takes the form $M_s\sim\mu_Bn(T_c/h)(T_c/E_F)\sim\mu_Bn\times10^{-5}$, where $h$ is the exchange field in the energy units, $n$ is the electron concentration, and $\mu_B$ is the Bohr magneton, $T_c$ is a critical temperature of an infinite superconductor and $E_F$ is its Fermi energy\cite{Devizorova-PRB-19}.
The relative contribution of this induced magnetization to the magnetic field distribution in S/F structures becomes small if the magnetism originates from the localized spins which is typically of the order of the Bohr magneton $\mu_B$ per atom but can be important if the magnetization in F layers is associated with the itinerant electrons.
In the latter case of significant spin polarization near the S/F boundary the corresponding magnetic moment can partially compensate the magnetic moment of the F layers. Assuming both the coherence length and the F layer thickness to be much smaller than the London penetration depth one can easily take into account this partial magnetic moment compensation replacing the bare magnetization $M_0$ in our expressions by the effective magnetization  averaged over the region of the F layer and the adjacent S layer region of the thickness $\sim\xi_s$.

The paper is organized as follows.
In Sec.~\ref{Sec_Model} we discuss the basic equations which allow to calculate the magnetic field profiles in multilayered S/F structures. In Sec.~\ref{Sec_Infinite} we analyze the electromagnetic proximity effect in the infinite S/F superlattice and calculate the spatial distribution of the induced magnetic field. In Sec.~\ref{Sec_Edges} we analyze the magnetic field profiles near the sample surface and calculate the characteristic decay lengths corresponding to different orientations between the sample surface, the S/F interfaces and the magnetic moment direction in the F layers. In Sec.~\ref{Sec_Finite} we calculate the magnetic field profiles for the S/F structures with the finite number of periods $N$ and compare the results with the recent experimental data on the neutron scattering in Nb/Gd superlattices. In Sec.~\ref{Sec_AntiF} we demonstrate the presence of strong EPE in S/F superlattices where the magnetic moments in the neighboring F layers have opposite direction. Finally, in Sec.~\ref{Sec_Concl} we summarize our results.

\section{Model}\label{Sec_Model}

\begin{figure}[b!]
    \includegraphics[width=0.8\linewidth]{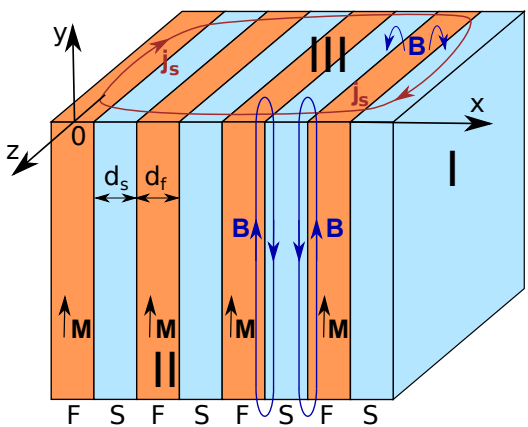}
    \caption{Sketch of S/F superlattice. The three types of surface with different orientation with respect to the S/F interfaces and the direction of magnetic moment in the F layers are indicated as I, II, and III. The directions of the magnetic field and superconducting currents are shown by the blue and red lines, respectively.
    }
    \label{Fig_ML_Scheme}
\end{figure}

We consider the multilayer S/F structure consisting of the identical S layers of the thickness $d_s\ll\lambda$ and the F layers of the thickness $d_f\ll\lambda$ (see Fig.~\ref{Fig_ML_Scheme}). The magnetization vectors ${\bf M}$ in all F layers are assumed to have the same magnitude $M_0$ and to be directed along the $y$ axis. Also we assume that the sample has the form of the brick with the faces oriented parallel or perpendicular to the layers and magnetization $\mathbf M$. They  can be classified into three types (which are marked as I, II and III in Fig.~\ref{Fig_ML_Scheme}) depending on their orientation with respect to the S/F interfaces and the direction of the magnetization ${\bf M}$ in the F layers, which is important for the future analysis.

To calculate the spatial profile of the magnetic field induced due to EPE we use the standard London approach assuming the local relation ${\bf j_s}({\bf r})=-(c/4\pi)\lambda^{-2}({\bf r}){\bf A}(\bf r)$ between the superconducting current ${\bf j}_s$ and the vector potential ${\bf A}$ which is typical for the systems in the dirty limit. The Maxwell equation for the magnetic field ${\bf B}={\rm rot}\,{\bf A}$ takes the form
\begin{equation}
{\rm rot}\,{\rm rot}\,{\bf A}+\lambda^{-2}({\bf r})\,\mathbf A=4\pi\,{\rm rot}\,\mathbf{M}.
\label{Eq_L}
\end{equation}
Here we take into account that the total current contains two contributions ${\bf j}={\bf j}_s+{\bf j}_m$ where ${\bf j}_m=c\,{\rm rot}\,{\bf M}$ is the magnetization current flowing along the edges of the ferromagnetic layers. The magnetization is taken in the form $\mathbf M=M(\mathbf r)\mathbf y_0$ with $M(\mathbf r)=M_0$ inside the F layers and $M(\mathbf r)=0$ elsewhere.

The Cooper pairs penetrating the ferromagnet induce the superconducting correlations there. As a result, the screening parameter $\lambda^{-2}$ becomes nonzero inside the F layers. Our main results do not depend on the specific form of the $\lambda^{-2}(x)$ profile ($x$ is the coordinate axis across the layers with the origin in the center of the F layer) which is only assumed to be a periodic function with the period $d=d_s+d_f$. However, to obtain the quantitative results relevant for the specific S/F structure one may use, e.g., the Usadel theory\cite{} which allows the calculation of the function $\lambda^{-2}(x)$. As an illustration, let us consider the limiting case when the exchange field exceeds the superconductor critical temperature $h\gg T_c$ and the normal-state conductivity $\sigma_s$ of the S layers strongly exceeds the conductivity $\sigma_f$ of the F layers. Then from the well-known solution of the linearized Usadel equation with the rigid boundary conditions at the S/F interfaces\cite{Mironov-APL-18} we find that inside the S layers the London penetration depth $\lambda(x)=\lambda_s=\mathrm{const}$ while in the $n$-th F layer occupying the region $-d_f/2<(x-nd)<d_f/2$ the screening parameter takes the form (see Appendix.~\ref{Sec_App} for the calculation details)
\begin{equation}\label{L_SF_res}
\lambda^{-2}(x)=\frac{1}{\lambda_s^2}\mathrm{Re}\left[\frac{\cosh^2q(x-nd)}{\cosh^2\left(qd_f/2\right)}\right].
\end{equation}
Here $q=\sqrt{2i}/\xi_f$ and $\xi_f$ is the coherence length inside the ferromagnet. To provide an effective interaction of the superconducting correlations through the F layers  we assume $d_f\le\xi_f$ in this paper.

\section{Spontaneous magnetic field in S/F superlattices with large number of layers}\label{Sec_Infinite}

We start from the simplest case when all sizes of the S/F superlattice strongly exceed the London penetration depth $\lambda$ and, thus, the typical scale of the magnetic field variations. As a first step, let us calculate the distribution of the magnetic field far (at distances much larger then $\lambda$) from the sample surface where the superlattice can be considered as infinite.

To solve Eq.~(\ref{Eq_L}) deep in the bulk of the sample we may choose the vector potential $\mathbf A=A(x)\mathbf z_0$. Obviously, the validity of such an ansatz for ${\bf A}$ breaks down at distances $\lesssim \lambda$ from the sample surface. The later case will be considered in detail in Sec.~\ref{Sec_Edges}. Far from the surface Eq.~(\ref{Eq_L}) for the function $A(x)$ takes the form 
\begin{equation}
A_{xx}''(x)-\lambda^{-2}(x)A=-4\pi M_x'(x),
\label{Eq_L1D}
\end{equation}
Disregarding the edge effects we will search the spatially periodic solution $\tilde A(x)$ for the vector potential so that $\tilde A(x+d)=\tilde A(x)$ where $d=d_s+d_f$. It is convenient to perform the Fourier transform of all terms in Eq.~(\ref{Eq_L1D}):
\begin{align}\label{FExpand}
& \tilde A(x)=\sum_n \tilde A_n\exp(iknx),\quad k=2\pi/d,\\
& 4\pi M_x'(x)=\frac{8\pi iM_0}{d}\sum_n\sin\left(\frac{knd_f}{2}\right)\exp(iknx),\\
& \lambda^{-2}(x)=\sum_n L_n\exp (iknx).
\end{align}
For the specific case when the origin of the $x$ axis is chosen in the middle of the F layer the coefficients $L_n=L_{-n}$ are real [see the Appendix for the explicit expression for $L_n$ corresponding to the profile (\ref{L_SF_res})]. Assuming $d\ll \lambda$ we solve the Eq.~(\ref{Eq_L1D}) perturbatively keeping the terms up to the order $\sim O\left[(d/\lambda)^2\right]$. The resulting expressions for the coefficients $\tilde A_n$ read
\begin{align}
& \tilde A_0=0,\label{Eq_A0}\\
& \tilde A_{n\neq 0}=\frac{4iM_0}{kn^2}\left[\sin(kns)-
\sum_{m\neq 0}\frac{L_{n-m}}{k^2m^2}\sin(kms)\right],\label{Eq_An}
\end{align}
where $s=d_f/2$.

In Eq.~(\ref{Eq_An}) the dominant contribution comes from the first term which does not depend on $\lambda$. This vector potential corresponds to $B=4\pi M_0d_s/d$ inside each ferromagnetic layer and $B=-4\pi M_0d_f/d$ in a superconducting layer. Remarkably, the magnetic field averaged over the S/F period of the superlattice $\langle B\rangle=0$. This regime is analogous to the one realized in the ferromagnetic superconductors where the Meissner currents fully compensate the magnetic field produced by magnetization. The fact of such total compensation does not depend on the specific magnitude of the London penetration depth provided we consider the region far from the sample surface. 

The second term in Eq.~(\ref{Eq_An}) is a correction describing the inhomogeneity of the Meissner screening currents at the length-scale of the order of $d\ll \lambda$. The resulting profile of the magnetic field accounting for this correction takes the form
\begin{equation}\label{EQ_BFull}
\tilde B(x)=4\pi\left[M(x)-\left< M\right>\right]+\sum_{m,n\neq0}\frac{4M_0L_{n-m}}{k^2m^2n}e^{iknx}\sin(kms),
\end{equation}
where $\langle M\rangle =M_0d_f/d$ is the magnetization averaged over the S/F structure period and $L_n$ are the Fourier harmonics of $\lambda^{-2}(x)$. Although the effects coming from the inhomogeneity of the Meissner currents are small for the structures with $d\ll\lambda$ they can become significant in the systems with $d\sim\lambda$ where they may result in the smoothing of the meander-like profiles of the magnetic field inside the superlattice.

\section{Edge effects}\label{Sec_Edges}

\begin{figure}[b!]
    \includegraphics[width=0.8\linewidth]{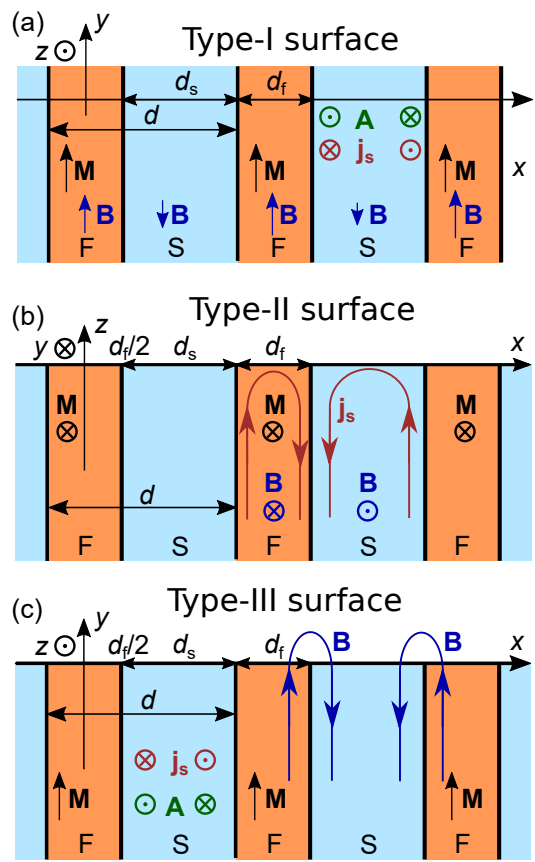}
    \caption{Schematic profiles of the magnetic field and superconducting currents near the three types of the sample surface.}
    \label{Fig_ML_Edges}
\end{figure}

In this section we take into account the finite dimensions of the S/F superlattice and analyze the profiles of the magnetic field near the sample surface of the three types shown in Fig.~\ref{Fig_ML_Edges}. Although deep in the bulk of the sample the average magnetic field vanishes (see Sec.~\ref{Sec_Infinite}) near its surface it becomes non-zero and its decay length depends on the type of the surface.

We start from the case of the type-I surface which is parallel to the S/F planes (Fig.~\ref{Fig_ML_Edges}a).
In this case the magnetic field and the corresponding vector potential are directed along the $y$ and $z$ axes, respectively, so that
\begin{equation}
\mathbf B=B(x)\mathbf y_0,\,\,\mathbf A=A(x)\mathbf z_0.
\end{equation}
Assuming that the total superlattice thickness strongly exceeds $\lambda$ we consider the semi-infinite structure with alternating S and F layers occupying the region $x>-d_f/2$ and a vacuum at $x<-d_f/2$. Inside the S/F lattice the vector potential satisfies Eq.~(\ref{Eq_L1D}). Searching the solution in the form
\begin{equation}
A(x)=\tilde A(x)+\sum\limits_{n} A_n\exp[i(kn+q)x]
\end{equation}
and performing the Fourier transformation we get:
\begin{equation}
(kn+q)^2A_n+\sum\limits_{m} L_{n-m}A_m=0.
\end{equation}
This system has a non-trivial solution only provided
\begin{equation}
q=i\sqrt{L_0}\left(1-\sum_{n>0}\frac{|L_n|^2}{L_0k^2n^2}\right).
\end{equation}
Here we take into account that $L_{-n}=L_n$. The value of $q$ is purely imaginary and the chosen sign of $q$ corresponds to the solution decaying towards the bulk of the sample. In the leading order over the small parameter $(d/\lambda_s)$ the characteristic decay length is $\lambda_0=L_0^{-1/2}=\langle\lambda(x)^{-2}\rangle^{-1/2}$. 

The amplitudes $A_n$ are defined by the boundary condition reflecting the continuity of the magnetic field at the outer boundaries of the sample:
\begin{equation}
A'(-d_f/2)=B_{ext}+4\pi M\approx A_h'(-d_f/2)+iqA_0,
\end{equation}
\begin{equation}
A_0=-\lambda_0\left(B+4\pi M_0\frac{d_f}{d}\right),
\end{equation}
where $B_{ext}$ is the external magnetic field, which is assumed to be directed along the $y$ axis. The corresponding solution for the magnetic field takes the form:
\begin{equation}
B(x)=\tilde B(x)+(B_{ext}+4\pi M_0d_f/d)\exp(-x/\lambda_0)
\end{equation}
Thus, near the type-I surface the magnetic field profile can be well-approximated by $B(x)=B_{ext}+4\pi M(x)$ while far from the surface the magnetic field profile is described by Eq.~(\ref{EQ_BFull}). 

Now we turn to the analysis of the case of the type-II surface (Fig.~\ref{Fig_ML_Edges}b). The S/F structure is assumed to occupy the half-space $z<0$. In this case the magnetic field is directed along the $y$ axis while the vector potential has two components in the $xz$ plane:
\begin{equation}
\mathbf B=B(x,z)\mathbf y_0,\,\,\mathbf A=A_z(x,z)\mathbf z_0+A_x(x,z)\mathbf x_0
\end{equation}
Far from the structure surface, i.e. at $y\rightarrow-\infty$, the superconducting current ${\bf j}_s$ and ${\bf A}$ have only one $z$ component.
However, in the vicinity of the surface the $x$ component should appear to guarantee the current continuity condition $\mathrm{div}\,\mathbf j_s=0$.

In this geometry it is more convenient to analyze the equation for the magnetic field $\mathbf H(\mathbf r)=\mathbf B(\mathbf r)-4\pi\mathbf M(\mathbf r)$ instead of equations for $\mathbf B(\mathbf r)$ or ${\bf A}({\bf r})$. 
Inside the S/F structure ${\bf H}$ satisfies the London type equation
\begin{equation}
-\mathrm{rot}\,\left(\lambda^2\mathrm{rot}\,\mathbf H\right)-\mathbf H=4\pi\mathbf M,
\end{equation}
and the boundary condition
\begin{equation}
H_y(z=0)=B_{ext}.
\end{equation}
We neglect a small oscillating component of $H$
and apply a perturbation approach with a small parameter $(d/\lambda)$.
In the first order it gives the following expression for $H_y$:
\begin{equation}\label{Eq_Type2_decay}
H_y=-4\pi\langle M\rangle+(B_{ext}+4\pi\langle M\rangle)\exp(z/\sqrt{\langle\lambda^2\rangle}).\\
\end{equation}
The corresponding magnetic field inside the sample can be restored by adding the F-layers magnetization field
\begin{equation}
    B_y(x,y)=4\pi M(x)+H_y(x,y),
\end{equation}
The type-II surface does not generate stray field so that the magnetic field is equal to the external field, i.e. $B_y=B_{ext}$. Remarkably, from Eq.~(\ref{Eq_Type2_decay}) one sees that the typical scale characterizing the magnetic field decay for the type-II surface differs from the one previously obtained for the type-I surface. Specifically, for the type-II surface it is equal to $\langle\lambda^2\rangle^{1/2}$.

Finally, we turn to the case of the type-III surface. We assume that the S/F structure occupies the half-space $y<0$ (Fig.~\ref{Fig_ML_Edges}c).
In contrast to the previous cases, the S/F lattice with the type-III boundary produces the nonzero stray magnetic fields outside the sample. The corresponding magnetic field near the type-III structure surface lies in the $xy$ plane while vector potential and superconducting current ${\bf j}_s$ have only the $z$-component:
\begin{equation}
\mathbf B=B_y(x,y)\mathbf y_0+B_x(x,y)\mathbf x_0,\,\,\mathbf A=A(x,y)\mathbf z_0.
\end{equation}
Applying the Fourier transformation to the $A(x,y)$ profile
\begin{equation}
A(x,y)=\sum\limits_n A_n(y)\exp(iknx)
\end{equation}
we obtain the following London equation:
\begin{equation}\label{Eq_Type3_London}
\frac{\partial^2 A_n}{\partial y^2}-(kn)^2A_n=\Theta(-y)\left[\sum_mL_{n-m}A_m+
4ikM_0\sin(kns)\right],
\end{equation}
where $\Theta(y)$ is the Heaviside step function. To match the solutions inside and outside the sample one needs to impose the boundary condition, which requires the continuity of $B_y$ at the sample surface. Then neglecting the small contributions $\sim O\left[(d/\lambda_s)^2\right]$ the solution of Eq.~(\ref{Eq_Type3_London}) takes the form:
\begin{align}
& A_0=0,\\
& A_n=-\frac{2iM_0d}{\pi n^2}\sin(kns)\left[\Theta(-y)+\frac{\mathrm{sgn}(y)}{2}\exp(-k|ny|)\right].
\end{align}
The corresponding magnetic field reads:
\begin{align}
& B_{yn}(x)=\frac{4M_0}{n}\sin(kns)\left[\Theta(-y)+\mathrm{sgn}(y)\frac{1}{2}\exp(-k|ny|)\right]\\
& B_{xn}(x)=\frac{2iM_0}{n}\sin(kns)\exp(-k|ny|).
\end{align}
Deep inside the sample ($y\rightarrow-\infty$) the magnetic field approaches  Eq.~(\ref{EQ_BFull}) while outside the sample the stray magnetic field decays at the scale of the order of $d$ so that far from the surface (at $y\rightarrow+\infty$) $B\to 0$. At the same time, near the type-III surface in the region of the thickness $d$ the magnetic field profile has the form schematically shown in Fig.~\ref{Fig_ML_Edges}c. In contrast to the previous cases, for an infinite type-III surface the average over the structure period $d$ superconducting currents are absent.

Thus, the magnetic field emerging due to the electromagnetic proximity effect near the surface of the S/F superlattice brick is sensitive to the relative orientation between the plane of the boundary, the direction of magnetization in the F layers and the interfaces between the S and F layers. 

\section{Giant electromagnetic proximity effect in S/F structures with finite number of layers}\label{Sec_Finite}

In this section we demonstrate that S/F superlattice with large but finite number of layers enables dramatic enhancement of the electromagnetic proximity effect as compared to the previously studied S/F bilayers \cite{Mironov-APL-18} and S/F/F spin valve structures \cite{Devizorova-PRB-19}. Specifically, in structures with S/F lattices placed on top of the semi-infinite superconductor the spontaneous magnetic field at the interface of the thick superconductor can reach the values of the order of the F layers magnetization $M_0$ while in S/F systems such field has the order of $(d_f/\lambda)^2M_0\sim 10^{-2}M_0$. To obtain such strong enhancement of the effect one needs to consider the S/F lattice of the total thickness of the order of $\lambda_s$ and $d\ll\lambda_s$.

\begin{figure}[b]
\includegraphics[width=0.95\linewidth]{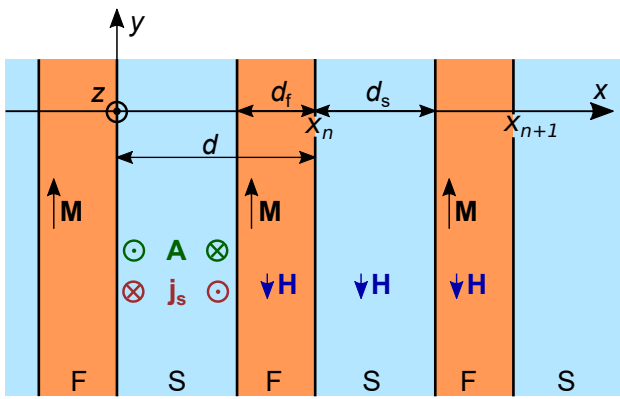}
\caption{Sketch illustrating the choice of the coordinate system used for the description of the S/F lattice with finite number of layers.}
\label{Fig_Matr_Scheme}
\end{figure}

In what follows we consider two types of structures: (i) S/F lattice placed on top of the bulk superconductor and (ii) isolated S/F superlattice surrounded by vacuum or insulator. We assume that the superlattice has the finite number $N$ of S/F periods and the layers have the infinite lateral size. To calculate the spatial profiles of the magnetic field in these structures we apply the well-known transfer matrix method. In particular, we solve the London equation inside each S/F period, thus, obtaining the relations between the values of the magnetic field $H$ and the normalized vector potential $a=A/\lambda_0$ at the opposite sides of S/F period (at $n$-th and $(n+1)$-th S/F interface, where $n$ enumerates the S/F periods). In this approach the difference between two types of structures described above is accounted for by different boundary conditions at the interface between the superlattice and bulk superconductor or vacuum.  

\begin{figure}[b]
\includegraphics[width=0.95\linewidth]{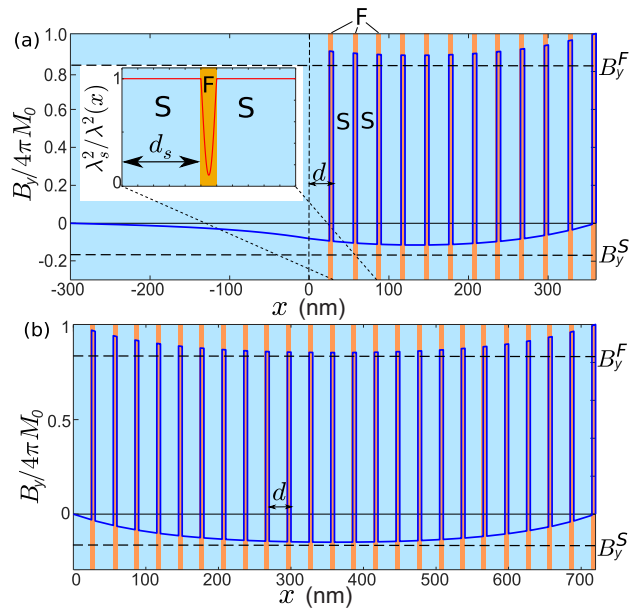}
\caption{The magnetic field profile $B(x)$ in (a) S/F superlattice with $N=12$ bilayers which is in contact with thick ($d_0=300$~nm) superconducting layer ($d_s=25$~nm, $d_f=5$~nm, $\lambda_s=120$~nm). The inset shows the profile $\lambda^{-2}(x)$ inside a single S/F period, $\xi_f=4$~nm. (b) A similar system with $N=24$ bilayers with vacuum or insulator on both outer interfaces.}
\label{fig_Matr}
\end{figure}

To solve the London equation for the $n$-th S/F bilayer we choose a coordinate system shown in Fig.~\ref{Fig_Matr_Scheme} so that the vector potential and the magnetic field depend only on the coordinate $x$ across the layers: $\mathbf A=a(x)\lambda_0\mathbf z_0$ and $\mathbf H=H(x)\mathbf y_0$. Taking ${\bf B}={\bf H}+4\pi {\bf M}$ we integrate the relation ${\rm rot}\mathbf{A}=\mathbf B$ and the London equation ${\rm rot}{\bf H}=-\lambda^{-2}(x)\mathbf A$ along the $x$ axis over the bilayer thickness $d$ and obtain the following integral equations for the functions $a(x)$ and $H(x)$:
\begin{align}
& a(x_{n+1})=a(x_n)-\int\nolimits_{x_n}^{x_{n+1}}\left[H(x')+4\pi M(x')\right]\lambda_0^{-1}dx',\label{Eq_LInt1}\\
& H(x_{n+1})=H(x_n)-\int\nolimits_{x_n}^{x_{n+1}}\lambda^{-2}(x')\lambda_0a(x')dx'.\label{Eq_LInt2}
\end{align}
Here the distribution $\lambda(x)$ is determined by Eq. (\ref{L_SF_res}),  $x_n$ and $x_{n+1}=x_n+d$ are the coordinates of the neighboring S/F interfaces chosen in a way that the $n$-th S layer occupies the region $x_n<x<x_n+d_s$ while the position of the $n$-th F layer corresponds to the region $x_n+d_s<x<x_{n+1}$ (see Fig.~\ref{Fig_Matr_Scheme}). These two linear equations establish the following relations between the values $a(x_n)$ and $H(x_n)$ with the values $a(x_{n+1})$ and $H(x_{n+1})$:
\begin{equation}\label{Eq_Diff}
v_{n+1}=\hat Cv_n+m,
\end{equation}
where
\begin{align}
& v_n=\left(\begin{array}{c}
a(x_n)\\H(x_n)
\end{array}\right),\,
\hat C=\left(\begin{array}{cc}
1+R & P\\
P & 1+S
\end{array}\right),\\
& m=4\pi M_0\left(\begin{array}{c}
-d_f/\lambda_0\\Q
\end{array}\right).
\end{align}
To get the values $R$, $S$, $P$, $Q$ we solve the integral equations (\ref{Eq_LInt1}) and (\ref{Eq_LInt2}) using the perturbation approach with the small parameter $d/\lambda_0\ll 1$. In the second order of the perturbation theory we obtain
\begin{align}
& R=\int_0^d\int_0^{x}\frac{dx'}{\lambda^2(x')}\,\,dx,\,
~~~P=-\frac{d}{\lambda_0},\\
& S=\int_0^d\frac{xdx}{\lambda^2(x)},\,
~~~Q=\int_{d_s}^d\frac{(x-d_s)dx}{\lambda^2(x)}.
\end{align}
In what follows we consider the superlattice consisting of $N$ spatial S/F periods so that the sample outer boundaries correspond to the planes $x_1=0$ and $x_{N+1}=Nd$.
The equation (\ref{Eq_Diff}) should be supplemented with the equations for the vectors $v_1$ and $v_{N+1}$ corresponding to the boundary conditions of the S/F lattice.

Let us start from the description of the giant electromagnetic proximity effect emerging in the systems where the S/F lattice is positioned on top of the bulk superconductor occupying the region $x<0$. In this case at $x<0$ the magnetic field and the corresponding vector potential exponentially decay towards the bulk of the superconductor: $B_y(x)=H(x)=H(0)\exp(x/\lambda_s)$ and $a(x)=a(0)\exp(x/\lambda_s)$. This gives us the following boundary condition coupling the functions $H(x)$ and $a(x)$ at $x=0$: $\lambda_sH(0)=\lambda_0a(0)$. At the same time, in the plane $x=Nd$ corresponding to the interface between the S/F lattice and vacuum one has $H(Nd)=0$ (here we assume for simplicity that there is no external magnetic field). As a result, in terms of the vector $v_n$ the above two conditions can be rewritten in the form
\begin{equation}\label{BC_2}
    v_1=\left(\begin{array}{c}
    a(0)\\\lambda_0a(0)/\lambda_s
\end{array}\right),
~~~v_{N+1}=\left(\begin{array}{c}
    a(Nd)\\0
    \end{array}\right)
\end{equation}

We introduce the auxiliary vector $w=(\hat I-\hat C)^{-1}m$ which satisfies the equation $w=\hat Cw+m$ (here $\hat I$ is the unit $2\times 2$ matrix).
Thus, from the Eq. (\ref{Eq_Diff}) the vectors $v_1$ and $v_{N+1}$ can be linked:
\begin{equation}
    v_{N+1}=w+\hat C^N(v_1-w).
    \label{Eq_matr1}
\end{equation}
The system of equations (\ref{Eq_matr1})-(\ref{BC_2}) enable to find the constant $a(0)$ and, thus, calculate the magnetic field amplitude $H(0)$.

To illustrate the dramatic enhancement of the electromagnetic proximity effect by the S/F lattice it is enough to keep only the terms up to the order $O(d/\lambda_0)$ so that one may put all the values $R$, $S$, and $Q$ equal to zero. Then 
\begin{equation}
    C^N\approx\left(\begin{array}{cc}
\cosh \psi & \sinh \psi\\
\sinh \psi & \cosh \psi
\end{array}\right),\,\,\,
\frac{w}{4\pi M_0}\approx\left(\begin{array}{c}
0\\d_f/d
\end{array}\right),
\end{equation}
where $\psi=Nd/\lambda_0$. Then for the magnetic field $B_y(0)$ we finally obtain:
\begin{equation}
    B_y(0)=4\pi M_0\frac{d_f}{d}\frac{\lambda_0(1-\cosh\psi)}
    {\lambda_0\cosh\psi+\lambda_s\sinh\psi}.
\end{equation}
In the limit $Nd\gg \lambda_0$ this expression reduces to
\begin{equation}\label{Eq_GiantB}
    B_y(0)=-4\pi M_0\frac{d_f\lambda_0}{d(\lambda_s+\lambda_0)}.
\end{equation}
One sees that in the case $d_f\sim d$ and $\lambda_s\sim\lambda_0$ the resulting magnetic field induced inside the bulk superconductor has the order of $M_0$. This results is in a sharp contrast with the case of the S/F bilayer where the spontaneous magnetic field $B_y\sim (d_f/\lambda_s)^2M_0\ll M_0$.
Thus, S/F lattices provide the way to increase the electromagnetic proximity effect by at least two order of magnitude. In Fig.~\ref{fig_Matr} we show the profile of the magnetic field for the specific S/F structure with $N=12$ periods on top of the bulk superconductor (all the system parameters are indicated in the figure caption).
The spatial profile of $\lambda^{-2}(x)$ distribution was found using the Eq. (\ref{L_SF_res}) for $\xi_f=4$~nm  (shown in the Fig.~\ref{fig_Matr} inset).
Note that the profile $\lambda^{-2}(x)$ in the $N$-th layer differs from the ones in all other F layers since this layer contacts with vacuum instead of a superconductor.
As a result, the matrix $\hat C$ for the $N$-th period should be modified in a way that the value $d_f/2$ in the Eq. (\ref{L_SF_res}) should be replaced with $d_f$.
This modification, however, does not change significantly the final result provided $N\gg1$.

Recently, the spatial profiles of the magnetic field inside the S/F superlattices were analyzed in the neutron scattering experiments\cite{Khaydukov-PRB-19}. To model the situation realized in this experiment we consider the S/F lattice which consists of $N=24$ periods and is surrounded by a vacuum so that $B_y(0)=B_y(Nd)=0$. Solving numerically Eq.~(\ref{Eq_matr1}) with the appropriate boundary conditions we calculate the spatial distribution of the magnetic field $B_y(x)$ inside the structure. We choose the following parameters of the lattice: $d_s=25$~nm, $d_f=5$~nm, $\lambda_s=120$~nm and $\xi_f=4$~nm. The resulting $B_y(x)$ distribution is shown in Fig.~\ref{fig_Matr}(b). Clearly, in the central part of the structure (far from the sample surface) the magnetic field profile approaches the one relevant for the infinite lattices described in Sec.~\ref{Sec_Infinite}: the magnetic field averaged over the structure period is equal to zero so that in the F layers $B_y^F\sim 4\pi M_0 d_s/d$ while in the S layers $B_y^S\sim -4\pi M_0 d_f/d$, (these values are marked with dash lines in Fig.~\ref{fig_Matr}).
It is important to note that  in Ref.~\onlinecite{Khaydukov-PRB-19} the magnetic field distribution was restored without taking into account the EPE effect.
Therefore, the $B(x)$ distribution in Ref.~\onlinecite{Khaydukov-PRB-19} qualitatively differs from the one shown in Fig.~\ref{fig_Matr}(b).

\begin{figure}[b]
\includegraphics[width=0.95\linewidth]{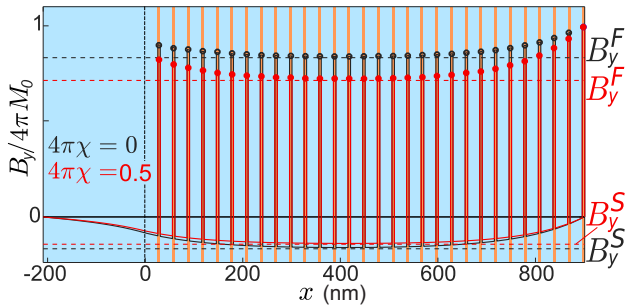}
\caption{ $B(x)$ distribution in a system with 30 bilayers ($d_s=25$nm, $d_f=5$ nm, $\lambda_s=120$~nm) on a thick ($d_0=210$ nm) superconducting layer for for fixed magnetization of the F layers ($\chi=0$, black line) and for variable magnetization ($M=\tilde M_0+4\pi\chi B$, $4\pi\chi=0.5$, red line). The values $B_y^S$ and $B_y^F$ for the two cases are shown by red and black lines, correspondingly. The open and filled circles shows the magnetic field in the F layer for the cases $\chi=0$ and $4\pi\chi=0.5$.
}
\label{fig_Variable}
\end{figure}

Comparing our results with the experimental data one can argue that in the experimentally realized structures the magnetic moment of each F layer in S/F lattice is not necessarily fixed and can vary self-consistently with the variation of the spontaneous field produced by all other layers due to the electromagnetic proximity effect. However, as we show below, this does not change our conclusions qualitatively only reducing the value of the magnetic field in all layers. Without going into the detail of the microscopic mechanisms beyond the magnetic response of the F layers we consider a simplified model which assumes the linear relation between the magnetization $M$ in the F layers and the local magnetic field $B$ characterized by the susceptibility $\chi=\partial M/\partial B$ ($4\pi\chi<1$). We assume that both $M$ and $B$ are oriented along $y$-axis, so that the linear relation takes the form
\begin{equation}
    M=\tilde M_0+\chi B,
    \label{MvsB}
\end{equation}
where $\tilde M_0$ is a constant which can be expressed through the magnetization $M_F$ of the isolated ferromagnet in the absence of the external magnetic field as $\tilde M_0=M_F(1-4\pi\chi)$. 
Substituting the relation (\ref{MvsB}) into the integral equations (\ref{Eq_LInt1},\ref{Eq_LInt2}) we get modified expressions of the matrix $\hat C$ and the vector $m$.
It results in the renormalization of the effective penetration depth which takes the form
\begin{equation}\label{Lambda_eff}
    \tilde\lambda=\left\{\int\frac{1}{\left[1-4\pi\chi(x)\right]\lambda^2(x)}dx\right\}^{-\frac{1}{2}}.
\end{equation}
Here $\chi(x)$ is constant in the F layers and equal to zero inside the S layers.
For the case of the uniform profile $\lambda(x)=\lambda_0$ relevant, e.g., for the structures with $d_f\ll\xi_f$ Eq.~(\ref{Lambda_eff}) reduces to
\begin{equation}
    \tilde\lambda=\lambda_0\sqrt\alpha,\,\,{\rm where}~
\alpha=\frac{1-4\pi\chi}{1-4\pi\chi d_s/d}<1.
\end{equation}

To calculate the profile of the magnetic field inside the S/F lattice one needs to account the relation (\ref{MvsB}) self-consistently. The magnetization inside the ferromagnets becomes dependent on the number of the F layer. To analyze how the dependence $M(B)$ affects the magnetic field distribution we compare two cases $\chi=0$ and $4\pi\chi=0.5$ considering the S/F lattice with $N=30$ periods on top of a 210-nm thick superconductor (see Fig.~\ref{fig_Variable}). The linear dependence between magnetization and the field $B$ results in the decrease of the magnetization inside the F layers as compared to the case of fixed magnetization which comes from the opposite directions between the magnetization ${\bf M}_0$ and the local magnetic field intensity ${\bf H}$. Far from the sample surface the magnetic field in the S and F layers ($B_y^S$ and $B_y^F$, respectively) takes the values
\begin{align}
    & B_s=-4\pi M_Fd_f\alpha/d,\\
    & B_f=4\pi M_Fd_s\alpha/d.
\end{align}
Thus, the dependence of the magnetization on the local magnetic field results only in the small damping of the magnetic field in all layers (see the factor $\alpha<1$) without any qualitative changes in the described phenomena.

\section{Superlattices with antiferromagnetic ordering}\label{Sec_AntiF}

\begin{figure}[b]
\includegraphics[width=0.95\linewidth]{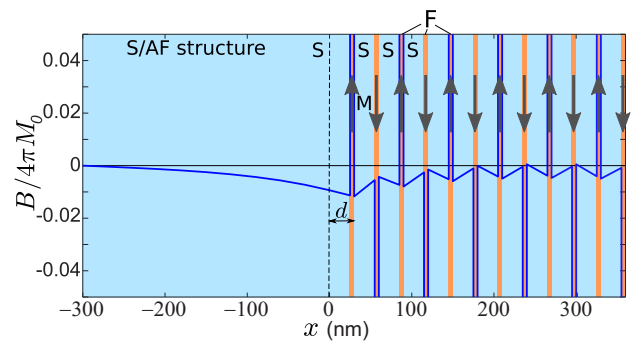}
\caption{Distribution of magnetic field $B_y(x)$ in a multilayer structure with antiferromagnetic ordering ($d_s=25$nm, $d_f=5$ nm, $\lambda_s=120$~nm) which is placed on the thick ($d_0=300$~nm) superconductor. We show the range $-0.2\pi M_0<B<0.2\pi M_0$ to visualize the distribution $B(x)$ in superconducting layers.
}
\label{fig_AF}
\end{figure}

Interestingly, the electromagnetic proximity effect can produce stray magnetic field even when the magnetic moments in two neighboring F layers of the S/F superlattices have the opposite directions (we will refer to such situation as to the case of antiferromagnetic ordering). For such type of structures one can naively expect that the electromagnetic proximity effect should be small since the average magnetization is zero. However, it is not the case: the S/F lattice positioned on top of the bulk superconductor induces spontaneous magnetic field of the order of $(d_f/\lambda)M_0$ in this superconductor which is larger than in a single S/F bilayer.

To demonstrate the origin of this effect we consider the S/F lattice where the projection of magnetization of the $n$-th F layer to the $y$ axis is equal to $M_0$ for even $n$ and $-M_0$ for odd $n$. The thickness of all F layers is assumed to be the same and equal to $d_f$. To calculate the magnetic field profiles we use the relations (\ref{Eq_LInt1}) and (\ref{Eq_LInt2}) which are valid for the structures under consideration. Following the calculation procedure, which is similar to the one used in Sec.~\ref{Sec_Finite}, we obtain the equation:
\begin{equation}\label{Eq_AF}
    v_{n+1}=\hat Cv_n+(-1)^{n-1}m.
\end{equation}
and the boundary conditions (\ref{BC_2}).
Note that the spatial period of the lattice with alternating direction of the magnetic moments is equal to $2d$. For an even layer number $n$ the equations takes the following form:
\begin{align}
    & v_{n+1}=\hat Cv_n+m,\\
    & v_{n+2}=\hat Cv_{n+1}-m=\hat C^2v_n+(\hat C-\hat I)m.
\end{align}
The further procedure is analogous to the ferromagnet ordering. We introduce an auxiliary vector $w=-(\hat C+\hat I)^{-1}m$ which satisfies the equation $w=\hat C^2w+(\hat C-\hat I)m$.
Then solving Eq.~(\ref{Eq_matr1}) with the new expression for the vector $w$ we find that the solution coincides with Eq.~(\ref{Eq_GiantB}) where one should replace $d_f/d$ with $-d_f/2\lambda_0$. In the case of thick anti-ferromagnetically ordered S/F lattice ($Nd\gg\lambda$) placed on top of the thick superconducting layer occupying the region $x<0$ the value of spontaneous magnetic field at the interface between the S/F lattice and the bulk superconductor is equal to
\begin{equation}
    B_y(0)=2\pi M_0\frac{d_f}{\lambda_s+\lambda_0}.
\end{equation}
Surprisingly, this field exceeds the spontaneous field induced by a single ferromagnet with $d_f\sim\xi_f$ (the later field has the order of $(d_f/\lambda)^2M_0$)\cite{Mironov-APL-18}. The numerical solution of Eq.~(\ref{Eq_AF}) for the case of the anti-ferromagnetic ordering between the neighboring F layers in the S/F lattice confirms the above result. The corresponding profiles of the magnetic field are shown in Fig.~\ref{fig_AF}. Thus, even in the case when the average magnetization of the S/F multilayered structure is zero the electromagnetic proximity effect induced by this magnetization can remain significant.

\section{Conclusion}\label{Sec_Concl}


To sum up, we demonstrate that the electromagnetic proximity effect in superlattices consisting of alternating ferromagnetic and superconducting layers can be strongly enhanced for the case of thick superlattices.
Besides that, the S/F superlattice can induce the spontaneous magnetic field in the adjacent bulk superconductor which is of the order of the magnetization field $M_0$ inside the ferromagnetic layers (while in S/F bilayers this field is damped by the factor $\sim (d_f/\lambda)^2\sim 10^{-2}$). 
In the S/F lattice far from the sample surface (i.e. at distance much larger than $\lambda$) the magnetic field $B$ averaged over the spatial period of the structure is zero (similar to the situation in the ferromagnetic superconductors) so that inside the F layers the magnetic field is equal to $B_F=4\pi M_0 d_s/(d_s+d_f)$ while in the S layers this field has the opposite direction and is equal to $B_S=-4\pi M_0 d_f/(d_s+d_f)$. Such full compensation of the magnetization field inside the F layers by the Meissner screening current is crucial for the adequate interpretation of the upcoming data in the neutron or muon scattering experiments. At the same time, near the sample surface the profiles of the magnetic field are substantially modified. In particular, the characteristic length of the magnetic field decay as well as the ability to generate the stray magnetic fields in the outer space of the sample strongly depends on the relative orientation between the plane of the surface, the plane of the layers and the magnetization direction. This effect may have a significant influence on the spontaneous fields generated in the experimentally realizable finite-size samples. Finally, we show that in S/F lattices with the opposite directions of the magnetic moments in the neighboring F layers the electromagnetic proximity effect is responsible for the generation of spontaneous magnetic field with the magnitude of the order of $(d_f/\lambda)M_0$. Remarkably, these fields appear to be much larger than the fields generated in S/F bilayers. 

Note that that the spontaneous magnetic fields induced by S/F superlattices can be larger than the superconducting lower critical field $H_{c1}$ which suggests the possible generation of Abrikosov vortices if the structure thickness exceeds $\lambda$. The back-action of vortices on the superconducting condensate responsible for the EPE should produce a variety of non-uniform magnetic and superconducting states similar to the ones in ferromagnetic superconductors. 

In addition, S/F superlattices may significantly enhance the long-range magnetic interactions between magnetic moments of ferromagnets recently predicted for the F/S/F sandwiches \cite{Devizorova-PRB-19} and, thus, become a perfect platform for the engineering of artificial complex magnetic interactions in S/F heterostructes. Also it is interesting to analyze the Josephson transport through S/F lattices since the spontaneous magnetic fields generated due to the proximity effect may significantly influence the behavior of the critical current in external magnetic field. Since the described phenomena do not require specific conditions and are expected to arise in typical experimentally realizable S/F structures we hope that our predictions could be verified already in the near future.

\section*{Acknowledgements}

The calculation of edge effects, and analysis of magnetic field distribution in structures with a finite number of layers with ferromagnetic ordering was supported by the Russian Science Foundation (Grant No. 20-12-00053). The calculations of the EPE effect in S/F structures with antiferromagnetic ordering was supported by the French ANR OPTOFLUXONICS and EU COST CA16218 Nanocohybri.

\appendix

\section{Calculation of the London penetration depth profile}\label{Sec_App}

To calculate the spatial profile of the magnetic screening parameter $\lambda^{-2}(x)$ we solve the linearized Usadel equation and follow the procedure used in Ref.~\onlinecite{Mironov-APL-18}. The main difference from the Ref.~\onlinecite{Mironov-APL-18} is that in our case the F layer is put between two superconducting layers, so that we impose the rigid boundary conditions on both S/F interfaces. The Eq. (\ref{L_SF_res}) gives us a resulting $\lambda^{-2}(x)$ distribution (\ref{L_SF_res}) inside the F layer. The explicit expressions for the Fourier harmonics $L_n$ of $\lambda^{-2}(x)$ can be calculated straightforwardly and take the form
\begin{multline}
L_n=\frac{1}{kd\lambda_s^2}\left[\sin\left(\frac{knd}{2}\right)-\sin(kns) \right]+\frac{2}{d\lambda_s^2}\left[
\frac{\sin(kns)}{k\cosh^2(qns)}+\right.\\
\left.+\frac{q\sinh(2qns)\cos(kns)+k\sin(kns)\cosh(2qns)}{2(4q^2-k^2n^2)\cosh^2(qns)}+c.c.\right].
\end{multline}

\end{document}